\documentclass[twocolumn,showpacs,preprintnumbers,amsmath,amssymb]{revtex4}

\usepackage{graphicx}
\usepackage{dcolumn}
\usepackage{bm}

\newcommand{\ndt}{\noindent}

\begin{document}



%
%
\author{
{\rm The MACRO Collaboration} \\
\nobreak\bigskip\nobreak
\pretolerance=10000
M.~Ambrosio$^{12}$, 
R.~Antolini$^{7}$, 
A.~Baldini$^{13}$, 
G.~C.~Barbarino$^{12}$, 
B.~C.~Barish$^{4}$, 
G.~Battistoni$^{6,a}$, 
Y.~Becherini$^{2}$,
R.~Bellotti$^{1}$, 
C.~Bemporad$^{13}$, 
P.~Bernardini$^{10}$
% ...
\nobreak\bigskip\nobreak
}

\vspace{0.25in}%

\affiliation{
\footnotesize
1. Dipartimento di Fisica dell'Universit\`a  di Bari and INFN, 70126 Bari, 
 Italy \\
2. Dipartimento di Fisica dell'Universit\`a  di Bologna and INFN, 40126 
Bologna, Italy \\
3. Physics Department, Boston University, Boston, MA 02215, USA \\
4. California Institute of Technology, Pasadena, CA 91125, USA \\
5. Department of Physics, Drexel University, Philadelphia, PA 19104, USA \\
6. Laboratori Nazionali di Frascati dell'INFN, 00044 Frascati (Roma), Italy \\
7. Laboratori Nazionali del Gran Sasso dell'INFN, 67010 Assergi (L'Aquila), 
 Italy \\
8. Depts. of Physics and of Astronomy, Indiana University, Bloomington, IN 
47405, USA \\
9. Dipartimento di Fisica dell'Universit\`a  dell'Aquila and INFN, 67100 
L'Aquila, Italy\\
10. Dipartimento di Fisica dell'Universit\`a  di Lecce and INFN, 73100 Lecce, 
 Italy \\
11. Department of Physics, University of Michigan, Ann Arbor, MI 48109, USA \\
12. Dipartimento di Fisica dell'Universit\`a  di Napoli and INFN, 80125 
Napoli, Italy \\
13. Dipartimento di Fisica dell'Universit\`a  di Pisa and INFN, 56010 Pisa, 
Italy \\
14. Dipartimento di Fisica dell'Universit\`a  di Roma "La Sapienza" and 
INFN, 00185 Roma, Italy \\
15. Physics Department, Texas A\&M University, College Station, TX 77843, 
 USA \\
16. Dipartimento di Fisica Sperimentale dell'Universit\`a  di Torino and 
INFN, 10125 Torino, Italy \\
17. L.P.T.P, Faculty of Sciences, University Mohamed I, B.P. 524 Oujda, 
 Morocco \\
18. Dipartimento di Fisica dell'Universit\`a  di Roma Tre and INFN Sezione 
Roma Tre, 00146 Roma, Italy \\
$a$ Also INFN Milano, 20133 Milano, Italy \\
$b$ Also Istituto TESRE/CNR, 40129 Bologna, Italy \\
$c$ Also Department of Physics, Pennsylvania State University, University 
Park, PA 16801, USA \\
$d$ Also Universit\`a  di Trieste and INFN, 34100 Trieste, Italy \\
$e$ Also U. Minn. Duluth Physics Dept., Duluth, MN 55812 \\
$f$ Also Dept. of Physics, MIT, Cambridge, MA 02139 \\
$g$ Also Intervideo Inc., Torrance CA 90505 USA \\
$h$Also Dipartimento di Fisica dell'Universit\`a  della Calabria, Rende 
(Cosenza), Italy \\
$i$ Macalester College, Dept. of Physics and Astr., St. Paul, MN 55105 \\
$j$ Also Department of Physics, James Madison University, Harrisonburg, VA 
22807, USA \\
$k$ Also Institute for Space Sciences, 76900 Bucharest, Romania \\
$l$ Also Universit\`a  della Basilicata, 85100 Potenza, Italy \\
$m$ Also Dipartimento di Scienze e Tecnologie Avanzate, Universit\`a  del 
Piemonte Orientale, Alessandria, Italy \\
$n$ Also Resonance Photonics, Markham, Ontario, Canada\\
}


\date{\today}


\title{The Search for the Sidereal and Solar Diurnal Modulations
in the Total MACRO Muon Data Set}

%


\begin{abstract}                
We have analyzed 44.3M single muons collected by MACRO from 1991 through 2000
in 2,145 live days of operation.  We have searched for the solar diurnal,
apparent sidereal, and pseudo-sidereal modulation of the underground muon rate
by computing hourly deviations of the muon rate from 6 month averages.  We
find evidence for statistically significant modulations with the solar
diurnal and the sidereal periods.  The amplitudes of these modulations are $<
0.1\%$, and are at the limit of the detector statistics.  The pseudo-sidereal
modulation is not statistically significant.  

The solar diurnal modulation is due to the daily atmospheric
temperature variations at 20 km, the altitude of primary cosmic ray
interactions with the atmosphere; MACRO is the deepest experiment to
report this result.  The sidereal modulation is in addition to the
expected Compton-Getting modulation due to solar system motion
relative to the Local Standard of Rest; it represents motion of the
solar system with respect to the galactic cosmic rays toward the
Perseus spiral arm.
\end{abstract}

\pacs{95.85.Ry, 98.70.Sa}

%
\maketitle

\section{Introduction}              

MACRO is a deep underground experiment designed to search for monoples
and rare particles in the cosmic rays.  Because of its great depth and
sensitivity to high energy primaries, the apparatus also probes the
physics of cosmic rays with energies in excess of 10 TeV.  Here we
describe a search for the first harmonics of the sidereal and solar
diurnal modulations in the underground muon rate using the MACRO
detector.  Sidereal modulations result from the motion of the solar
system with respect to a locally isotropic population of cosmic rays.
For instance, a local source of cosmic rays could introduce a sidereal
wave into the underground muon rate.  Or, if the local density of halo
cosmic rays were sufficiently high, a sidereal wave would be imprinted
on the underground cosmic ray rate as the solar system moved through
them.  Solar diurnal modulations at energies greater than 10 TeV on
the other hand are mostly the result of meteorological effects at or
near the primary site of the cosmic ray interactions in the
atmosphere.

\section{The First Harmonic Sidereal and Solar Diurnal Modulations}

There are two primary processes that can lead to periodic modulations
in the underground muon rate.  One process, the Compton-Getting effect
\cite{compton}, is due to an observer's motion with respect to a
locally isotropic population of cosmic rays.  The modulation is
introduced as the Earth's rotation periodically turns an observer
toward and away from the direction of motion.  Since the Earth rotates
on its axis in a sidereal day, the Compton-Getting modulation is seen
when muons are binned in sidereal time.  The second process that leads
to a periodic modulation in the muon rate deep underground 
is the result of the
competition between pion decay and interaction in the upper
atmosphere.  As the atmosphere cools at night (or during the winter),
the density increases and the pions produced in cosmic ray collisions
with the atmosphere are fractionally more likely to interact than
decay than during the day (or summer) when it is warmer, leading to
a decrease in the muon rate.  The daily
modulation introduced by solar heating and cooling is seen when muons
are binned in solar-diurnal time; yearly or seasonal modulations are
seen when events are binned with a period of a (tropical) year
\cite{macro1,macronim}.

Underground experiments measure the muon rate $R$.  So a sensitive 
search for sidereal and solar diurnal modulations in MACRO is a 
search for periodic variations in the underground muon rate as
compared with the average rate, 

\begin{equation}
\Delta R/R_0 = \int\!\!\!\int (\Delta I/I_0) \,d \Omega\,d A ,
\end{equation}

\ndt where $\Delta R = R - R_0$, $R$ is the measured cosmic ray rate,
$R_0$ is the average cosmic ray rate, $\Delta I/I_0 = (I-I_0)/I_0$ is
the deviation from the average cosmic ray intensity, and the
integration is over the acceptance of the detector.

\subsection{The Compton-Getting Effect}

An observer moving with velocity ${\bf v}$ relative to the rest frame
of a cosmic-ray plasma will detect a deviation due to this motion from
the average cosmic-ray intensity, an effect first described by Compton
and Getting \cite{compton}.  If the cosmic rays have a differential power-law
energy spectrum, $E^{-\gamma}$, then the first harmonic of this
deviation is given by

\begin{eqnarray}
\label{cg}
  \Delta I(\psi)/I_0 & = & (2 + \gamma) (v/c)  \cos{\beta}
                        \cos{\delta} \nonumber \\
& & ~ \times \cos{2 \pi[(\lambda - \alpha)/24]},
\end{eqnarray}

\ndt where $\Delta I(\psi) = I(\psi) - I_0$; $I(\psi)$ is the cosmic
ray intensity in a direction with space angle $\psi$ between the
direction of the detector's maximum sensitivity, $(\beta, \lambda)$,
and the direction of motion, $(\delta, \alpha)$, in celestial
coordinates on the sky; $I_0$ is the average cosmic ray intensity; and
$(2 + \gamma) (v/c)$ is the first harmonic amplitude \cite{gleeson}.
In this expression, $\beta$ is the declination and $\lambda$ is the
right ascension (in hours) of maximum detector sensitivity, and
$\delta$ is the declination and $\alpha$ is the right ascension (in
hours) toward the direction of motion.  Since a cosmic-ray detector is
fixed on the rotating earth, it sweeps the sky with period equal to a
sidereal day.  Under these circumstances $\lambda = \lambda(t)$, and
the Compton-Getting signal is modulated with a period of a sidereal
day.

Surface detectors are typically most sensitive to the declination
crossing the local zenith and the right ascension crossing the meridian.  
For such a detector, $\beta$ is equal to the detector's
latitude and $\lambda$ is equal to
the local sidereal time.  The $\lambda$ response of the detector 
therefore implies that a Compton-Getting signal will be modulated in sidereal
time, $t_\star$, by the phase term

\begin{equation}
2 \pi [(\lambda - \alpha)/24] = 2 \pi (\tau_\star - \alpha/24) ,
\end{equation}

\ndt where the sidereal phase $\tau_\star = t_\star/24$. 

As a deep underground detector with angular-dependent acceptance and
overburden, MACRO's directional sensitivity $(\beta, \lambda)$ to
sidereal signals differs from the ideal surface detector described
above in two ways.  First, the detector response may be peaked away
from the meridian.  In this case, MACRO would detect the signal at a
sidereal hour displaced from its true right ascension.  The upper
panel of Fig.~1 shows a histogram of the hour angle of the arriving
muons; this histogram shows that the mean of the hour angle
distribution falls at $-7.7^\circ$ from the meridian, an offset that
translates to a correction of $-31^m$ to the sidereal hour of the
detected signal maximum.

Second, the detector response may be peaked away from the zenith.  For
a detector whose response peaks away from the zenith, the declination
of the detector's maximum sensitivity in eq.(\ref{cg}) is not simply
its latitude and $\beta$ needs to be replaced by an `effective
declination', $\beta_{eff}$.  It is evident from eq.(\ref{cg}) that
this replacement affects only the amplitude of the signal and not the
sidereal hour of its maximum.  For MACRO $\beta_{eff}$ can in
principle be determined by a peak in the declination distribution for
the muons used in the analysis.  The lower panel of Fig.~1 shows this
declination distribution.  The situation is clearly complicated by the
broad distribution with mulitple peaks at $15^\circ$, $42^\circ$, and
$64^\circ$. (MACRO's latitude = $42.4528^\circ$.) In the analysis
presented here, we use the mean of the distribution as the effective
declination, $\beta_{eff} = 37^\circ$.  It should be noted, however,
that a value of $\beta_{eff}$ between $15^\circ - 42^\circ$ only
alters the signal amplitude by less than 30\%.

Possible Compton-Getting signals with a sidereal period
might result from solar motion with respect to galactic
sources of cosmic rays \cite{cutler3} or from motion with respect to
halo cosmic rays residing in the galactic disk.  The effect of solar
system motions on the underground muon rate gives the modulated cosmic
ray intensity

\begin{eqnarray}
 \label{sidereal} 
  [\Delta I(\tau_\star)/I_0]_{CG}& = &(2 + \gamma)
  (v/c) \cos{\beta_{eff}} \cos{\delta} \nonumber \\
& & ~\times \cos{2 \pi (\tau_\star - \alpha/24)}.
\end{eqnarray}

\ndt Thus the Compton-Getting signal expected in the undergound
muon rate would have the form

\begin{equation}
 \label{rsid} 
 [ \Delta R(\tau_\star)/R_0 ]_{CG} = K_\star  
    \cos{2 \pi (\tau_\star - \alpha/24)}.
\end{equation}

\ndt One component of the Compton-Getting signal seen by MACRO in
sidereal time is due to the solar motion relative to the Local
Standard of Rest (LSR) \cite{cutler3}, the rest frame of the local
cosmic rays.  Adopting v = 16.5 km/s toward $l = 53^\circ$, $b = 25^\circ$ 
for this motion \cite{mihalas}, we compute $K_\star^{LSR} = 2.6\times 10^{-4}$ 
for the amplitude of this wave and  
$\alpha^{LSR} = 17.8^h$ for its right ascension.

Previous searches for this effect have given different results
depending on the energy of the primaries probed.  Underground muon
observatories and air shower arrays (E $>$ 10 TeV) find sidereal
modulations with amplitudes in the range $5 \times 10^{-4} - 10^{-3}$,
consistent with a drift velocity of a few hundred km/s \cite{muna}.
Shallow undergound muon telescopes and neutron monitors (E $<$ 500
GeV) observe statistically significant sidereal modulations that are
explained by solar wind effects \cite{hall}.  Underground muons
detected by MACRO have energies at the surface in excess of $E_\mu \ge
1.25$ TeV and are due to interactions of primaries whose energy
exceeds 10 TeV.  Therefore, the observed muon rates are unaffected by
the solar wind.

If cosmic ray muons deep underground are binned in local solar diurnal
time, $t_\odot$, there will be a Compton-Getting signal that is a
consequence of the earth's orbital motion through the local solar
system cosmic rays.  Since the muons have energies at the surface
$E_\mu > 1.25$ TeV that leave them unaffected by the solar wind, they
are locally isotropic when averaged over the solar diurnal period.
Ignoring the earth's orbital eccentricity, the Compton-Getting
modulation due to this orbital motion peaks near 6 am local time when
the detector is sensitive in a direction most nearly parallel to the
earth's orbital velocity, and the cosmic ray intensity is modulated in
solar diurnal time by the phase term
 
\begin{equation}
2 \pi (\lambda - \alpha_\odot) = 2 \pi (\tau_\odot - 6/24) ,
\end{equation}

\ndt where the solar diurnal phase $\tau_\odot = t_\odot/24$
\cite{cutler2}.  In this case,

\begin{eqnarray}
 \label{solar}
  [\Delta I(\tau_\odot)/I_0]_{CG} &=& (2 + \gamma) (v/c)
  \cos{\beta_{eff}} \nonumber \\
                  &&  ~\times     \cos{2 \pi (\tau_\odot - 0.25)}.
\end{eqnarray}

\ndt The Compton-Getting signal in solar
diurnal time would have the form

\begin{eqnarray}
 \label{rcgd} 
 [ \Delta R(\tau_\odot)/R_0 ]_{CG} &=& K_{CG}  
    \cos{2 \pi (\tau_\odot - 0.25)}
                  \nonumber \\
                  &=& 4.55 \times 10^{-4}\cos{\beta_{eff}}\nonumber \\
                 && ~\times    \cos{2 \pi (\tau_\odot - 0.25)},
\end{eqnarray}

\ndt where the first harmonic amplitude has been computed from the
earth's average orbital speed. 

\subsection{Atmospheric Effects}

In addition to the Compton-Getting modulation, atmospheric effects
modulate the muons in solar diurnal time, $[\Delta
I_(\tau_\odot)/I_0]_{atm}$.  This atmospheric modulation is the result
of density variations at the altitude of first interaction -- when
cold at night, the density at the primary interaction site is
fractionally higher than the daily average and pions are fractionally
more likely to interact than decay, thereby giving fewer muons
relative to the average underground muon rate.  During the day, the
reverse occurs.  The modulation of the underground muon rate in solar
diurnal time due to atmospheric density variations is given by

\begin{equation}
 \label{atmd}
 [\Delta R(\tau_\odot)/R_0 ]_\odot = K_\odot  
   \cos{ 2 \pi (\tau_\odot - \xi_\odot )},
\end{equation}

\ndt where $\xi_\odot$ is the phase of maximum modulation.
Atmospheric effects also modulate the underground muon intensity on a
yearly time scale.  This is the so-called `seasonal modulation',
$[\Delta I(\tau_{yr})/I_0]_{yr}$, where $\tau_{yr}$ is the yearly
phase.  If $N_\odot$ is the number of solar days elapsed since the
beginning of the year and $N_{sol}$ is the number of days in a
tropical year, then $\tau_{yr} = (N_\odot + t_\odot/24)/N_{sol}$ and
the seasonal modulation is proportional to $\cos{2 \pi (\tau_{yr} -
\xi_{yr} ) }$.  The seasonal modulation has the form

\begin{equation}
 \label{atm}
 [ \Delta R(\tau_\odot)/R_0 ]_{yr} = K_{yr}  
   \cos{ 2 \pi (\tau_{yr} - \xi_{yr} )}.
\end{equation}

\ndt In Fig.~2 we show for MACRO the percentage deviations of the
monthly average muon rate from the yearly average (solid circles), as
a measure of $ [ \Delta R(\tau_\odot)/R_0 ]_{yr}$, for the years
1991-1997.  The fit of eq.(\ref{atm}) to these data gives $K_{yr} =
0.011 \pm 0.0004$ and $\xi_{yr}
= 0.54 \pm 0.06$.  Superposed on the muon data are the deviations in
the mean monthly `effective temperature' of the atmosphere from the
yearly average, $\Delta T_{eff}/<T_{eff}>$ (open triangles), as
defined in \cite{macro1}.  We have repeated the analysis from
\cite{macro1} that computes the correlation of the mean monthly
deviation in the muon rate with the mean monthly deviation in the
effective temperature of the atmosphere,

\begin{equation}
 \label{corr}
   [ \Delta R(\tau_\odot)/R_0 ]_{yr}  = \alpha_T \Delta T_{eff}/<T_{eff}>.
\end{equation}

\ndt A regression analysis shows that $\alpha_T = 0.91 \pm 0.07$, or equivalently $0.42 \pm 0.03$ \%/K,
consistent with what was found in \cite{macro1}.  

\subsection{The Pseudo-Sidereal Modulation}

As first pointed out by Compton and Getting \cite{compton},
atmospheric effects can mimic a sidereal modulation if the yearly 
(seasonal) wave beats with a solar diurnal wave.  Under these circumstances,
the beats due to atmospheric effects have the form

\begin{eqnarray} 
 \label{beats}
 [2 \Delta R/R_0 ]_{beats}
         & =&   K_\odot K_{yr} \times \nonumber \\
          && \cos{ 2 \pi
          [(N_{sol}+1)\tau_{yr} - \xi_\odot - \xi_{yr} ] }  \nonumber \\ 
          && + K_\odot K_{yr} \times \nonumber \\
          && \cos{ 2 \pi [(N_{sol}-1)\tau_{yr} 
             - \xi_\odot + \xi_{yr} ] } \nonumber \\ 
          &=&  K_\odot K_{yr} \times \nonumber \\
           && \cos{ 2 \pi (\tau_\star 
             - \xi_\odot - \xi_{yr} ) } 
             \nonumber \\ 
          && +  K_\odot K_{yr} \times \nonumber \\
          && \cos{ 2 \pi (\tau_{pseudo} 
             - \xi_\odot + \xi_{yr} ) } . 
\end{eqnarray}

\ndt In this expression, the first modulation, with a period of
$(N_{sol}+1)$ solar days, represents a pseudo-sidereal modulation
resulting from atmospheric effects.  This modulation was named 
the `antisidereal' modulation by Farley and Storey \cite{farley}, and 
is often called the `antisidereal wave' in the literature.  
The second modulation, with a
period of $(N_{sol}-1)$, solar days represents a modulation in
unphysical `pseudo-sidereal' time, $t_{pseudo} = N_{sol} \times 
t_\odot/(N_{sol}-1) $, and is purely the result of the beating of the
yearly and daily solar modulations.

To correct for the pseudo-sidereal effect, the method described by
Farley and Storey \cite{farley} is often used.  In this method, the
amplitude of the pseudo-sidereal wave, which is equal to the amplitude of
the pseudo-sidereal wave, is determined by binning the data
in pseudo-sidereal time.  The phase of the pseudo-sidereal wave is
determined by reflecting the pseudo-sidereal wave vector about the solar
wave vector.  

\section{Data analysis}

This analysis presented here requires only the arrival time of each
muon.  In principle, the analysis is simple: the arrival time of each
muon is binned separately in histograms of sidereal, solar diurnal,
and pseudo-sidereal time; an accurate account of the live-time in each
histogram bin is accumulated; at the end, the number of events in each
bin is divided by the live-time in that bin to give a muon rate.  The
results are based upon separate analyses of the muon rate in each
histogram bin compared with the average rate for that histogram.

For the sidereal analysis, it must be emphasized that the results are
not based on tracking information, that is, the determination of the
right ascension and declination of each muon.  Here is the aim of the
sidereal analysis.  If the muon rate were completely uniform over the
celestial sphere, the detector would measure a uniform rate
underground as a function of sidereal hour.  If there were an excess
of muons from some particular celestial direction, MACRO would measure
an increase in the muon rate as the detector's maximum response swept
over that direction.  In the sidereal analysis, we are searching for
this excess.

  \subsection{Run/Event selection}

In this analysis we included data runs starting from the beginning of
MACRO data taking with 6 supermodules in November 1991 through May
2000.  (A complete description of the detector during its running is
given in \cite{macronim}.) A typical data run had a duration of 6-12
hours, and over 9,500 data runs were included in this analysis.  The
analysis proceeded by first dividing these individual data runs into
17 {\it run sets} of approximately 6 months duration during which the
detector acceptance remained constant.  We then filtered the data to
include only muons with single tracks in both views.  In this
analysis, we accept events that fire at least 3 planes in the lower
detector.  For events crossing only 3 planes, the trigger efficiency
is approximately 72\%; for events crossing more than 3 planes, the
trigger efficiency is $> 99\%$.  Once filtered, we compiled for each
{\it run set} a histogram of the single muon rate per hour for all
runs in that run set and then fitted a Gaussian to the resulting
distribution.

We implemented run cuts as follows.  A run was excluded from further analysis
if:

\begin{itemize}

  \item not all 6 supermodules were active, or
  
  \item the wire efficiencies were $<$ 90\% and/or the strip efficiencies
  were $<$ 80\% during the run, or
 
  \item the single muon rate for the run was $> 5 \sigma$ from the mean single muon
  rate for that run set.

\end{itemize}

\ndt For the efficiency cuts, wire and strip efficiencies were determined for
each run by computing the average number of wires and strips recording hits
for all single muons crossing 10 planes.

There are 44.3M muons in the total data set after data cuts.  The
total live-time for the included runs is 2,145 days.  The average
single muon rate in the total data set is 860.53 muons/hour/6
supermodules.

\subsection{Histograms of deviations from the mean solar diurnal,
apparent sidereal, and pseudo-sidereal muon rates}

We searched for the solar diurnal, apparent sidereal, and pseudo-sidereal
modulations as follows.  First, event histograms for each run were
constructed for the three periods by binning the arrival time of each
muon according to its: (1) local solar diurnal time at the Gran Sasso;
(2) local sidereal time; and (3) local pseudo-sidereal time.  The periods
of these modulations are: solar diurnal day = 86,400 seconds; sidereal
day = 86,164.09892 seconds \cite{allen}; the pseudo-sidereal day was
assumed to be longer than a solar day by the same fraction that a
sidereal day is shorter than a solar day, or 86,636.54693 seconds.
This pseudo-sdereal wave has zero phase at the autumnal equinox, when the
sidereal time, the solar time, and the pseudo-sidereal time are
coincident \citep{farley}.  In this analysis, the pseudo-sidereal time
was computed relative to the 1988 autumnal equinox, September 22,
1988, 19$^h$ 29$^m$ UT.  Second, the live time in each run was
similarly binned into three histograms.  The live time for a run was
computed as the difference between the arrival times of the first and
last muons.  The live time was added to the histograms from the time
of the first muon to the time of the last muon.  The rate histograms
for each run were then computed by dividing the contents of the
appropriate event histogram by the contents of its corresponding live
time histogram.

The rate histogram for each run was unpacked and the muon rate in each
bin, $r_i$, was compared to the mean muon rate for that {\it run set},
$\bar r_j$, and its fluctuation from the mean was computed as
$\delta r_{ij} = (r_i - \bar r_j)/\bar r_j$.  Each $\delta r_{ij}$ for
that run was then compared with the r.m.s. of the distribution of
fluctuations for the total data set and those $\delta r_{ij} > (3 \times
~ r.m.s.)$ were cut from the analysis.  This cut was made to exclude
the effect of the fluctuations found in the long, asymmetric
non-Gaussian tails in the distribution of fluctuations for the total
data set.  These outliers, which comprise much less than 1\% of the
data, mostly result from run starts, run stops, sudden data
spikes, and other nonphysical effects.  The results of the analysis
are relatively insensitive ($\sim 10\%$) to this cut over the range $(2.5-5)
\times~r.m.s.$.  The $\delta r_{ij}$ passing this cut were entered into
summary histograms for the three periods.

At the end of this process, after all runs were analyzed, there resulted a
set of three histograms with fluctuations from the mean muon rate binned in
solar diurnal time, sidereal time, and pseudo-sidereal time.  The contents
of each bin in these summary histrograms were then normalized by the number of
entries in that bin.  The resulting histograms of the normalized
fluctuations, $\Delta R/R_0$, for the solar diurnal, sidereal, and
pseudo-sidereal periods are shown in Figures~3, 5, and 6, respectively.

  \section{Phase Analysis}

We searched for the first harmonics of the solar diurnal, apparent
sidereal, and pseudo-isidereal modulations by fitting the histograms of
the fluctuations to the form:

\begin{equation}
\label{modulation}
  [\Delta R/R_0] = K \cos{ 2 \pi [(t - t_{max})/24]},
\end{equation}

\ndt where $\Delta R/R_0$ is the deviation from the average muon rate;
$K$ is the amplitude of the modulation; $t$ is the solar diurnal,
sidereal time, or pseudo-sidereal time; and $t_{max}$ is the time of
maximum.  The results of a $\chi^2$-fit to the solar diurnal and 
sidereal histograms are given in Table~1.
The fitted curves are superposed on the histograms in Figures~3 and 5.  
In Table~1, a correction of -0.3$^h$ to $t_{max}$ been applied
to the sidereal wave fit to account for MACRO's acceptance and
overburden, as described in \S II.A.  For the pseudo-sidereal histogram, the 
expected pseudo-sidereal modulation from eq.(\ref{beats}) is given in Table~1, with errors
computed by adding the errors in the yearly and solar diurnal modulations in quadrature.  
This expected wave has been fitted to the pseudo-sidereal histogram and the results shown in Figure~6.  

In addition, Table~1 gives the results of
fitting the null hypothesis to the three data sets -- the hypothesis
that no modulation is present in the deviations from the average muon
rate.  

\subsection{Solar diurnal modulation}

Table~1 shows that the solar diurnal modulation is a statistically
significant effect.  The origin of this modulation is most likely the
daily atmospheric temperature variations at approximately 20~km, the
altitude of the primary cosmic ray interactions with the atmosphere.

We have used the atmospheric temperature data from the European Centre
for Medium-Range Weather Forecasts (ECMWF) at 0h, 6h, 12h, and 18h UT,
derived using hydrostatic equations from satellite observations of the
upper atmosphere to test the hypothesis that the solar diurnal
modulation is due to temperature effects.  We retrieved temperatures
in a 25 point grid centered in the Gran Sasso region (41.5$^\circ$ N -
43.5$^\circ$ N; 12.5$^\circ$ E - 14.5$^\circ$ E) for the 7 atmospheric
depths used to calculate T$_{eff}$ \cite{macro1}.  The result is shown
in Fig.~4 superposed on the solar diurnal wave.  The errors are
obtained from the dispersion of the hourly data around their average
values for the 7 year period included in the analysis.  The excellent
agreement between the observed solar diurnal wave and that expected
from the daily temperature wave is apparent, thus offering strong
evidence for the meteorological origin of the solar diurnal wave in
the underground muon rate.  MACRO is the deepest experiment to report
this effect.

\subsection{Sidereal modulation}

Table~1 shows that the sidereal modulation is also a statistically
significant result.  The observed amplitude of $K = 8.1 \times
10^{-4}$ is significantly larger than expected for the Compton-Getting
effect due to the solar system motion with respect to the LSR, $K =
2.6 \times 10^{-4}$; the maximum observed phase of $t_{max} = 23.2^h$
is also significantly different from the expected $t_{max} = 17.8^h$.
This suggests that we have found another modulation in addition to
expected Compton-Getting modulation due to solar system motion.

\subsection{Pseudo-sidereal modulation}

As shown in Table~1, the fit of eq.(\ref{beats}) to the psuedo-sidereal
histogram has the same statistical significance as the null hypothesis.
This result is not unexpected.  Since
the significance of the solar diurnal and sidereal modulations are 
detections at or near the limit of the detector statistics,
the pseudo-sidereal modulation is too small for a statistically significant 
signal.

\section{Results}

Since the pseudo-sidereal modulation is of low statistical significance,
the correction to the apparent sidereal modulation for the
pseudo-sidereal modulation is not required.  After correcting the
sidereal modulation for the solar motion relative to the LSR, the
final value for the sidereal modulation of the underground muon rate
observed by MACRO has an amplitude $K_\star = (8.2 \pm 2.7)
\times 10^{-4}$ and a phase of maximum of $t_{max} = 0.4 \pm 1.1$ hr.

The comparison of the result found in this investigation with the
results of other experiments in the primary cosmic ray energy range
$E_p \ge 1$ ~TeV are shown in Fig.~7.  In the upper panel the
amplitudes of the modulations are shown.  To compare the results of
different experiments, the reported amplitudes have been corrected for
the effective declination of the experiment, c.f. eq(\ref{cg}) and to the
LSR \cite{cutler3, lee}.  Included in the amplitude is a systematic
error of 7\% to account for the uncertainty in $\beta_{eff}$.  In
the lower panel, the sidereal hour of the maximum amplitude of the
modulation, or the right ascension of the maximum signal, is compared
with results from other experiments.  Again, the phase has been
corrected to the LSR.  In this figure, the y-axis labels for 18$^h$ to
24$^h$ have been replaced by -6$^h$ to 0$^h$.  These figures clearly
show that the results found in this investigation are consistent with
other experimental determinations in the energy range $10^{12} - 3
\times 10^{14}$ eV, and that these results provide a bridge between
underground experiments and extensive air shower arrays.

The interpretation of the sidereal wave is not straightforward due to
the complicated nature of cosmic ray propagation through the galactic
magnetic field at particle energies in the range $10^{12}-10^{15}$ eV.
However, some understanding can come from mapping the observed
sidereal wave into galactic coordinates.  We have realized this
mapping by first recognizing that MACRO is most sensitive to the
declination equal to $\delta \sim \beta_{eff} \sim 37^\circ$, and
then tracing out the path of this declination in sidereal time.  Once
this path has been determined, we transform it into galactic
coordinates \cite{duffet}.  This path is shown in Fig.~8.  Onto this
path we superpose the sidereal wave as follows: positive wave
amplitudes are shown as crosses and negative wave amplitudes are shown
as open circles.  The position on this path of the maximum wave
amplitude is shown as a filled star.  To show the effect of the
largest possible variation in $\beta_{eff}$, the position of the
maximum wave amplitude for $\beta_{eff} = 15^\circ, 64^\circ$, the 
outlying peaks in the 
declination distribution in Fig.~1, are
shown as filled triangles.  In addition we show the direction toward
which the sun is moving as a result of differential galactic rotation
as a filled circle and the direction of the Perseus spiral arm as a
filled square.  This figure shows that the wave intensity is positive
over a wide range in galactic longitude, $l = 80^\circ - 150^\circ$,
when MACRO is pointing towards the galactic plane, and is negative
when MACRO is pointing to high galactic latitudes.  This pattern is
typical of many detectors in the northern hemisphere \cite{naga,
gombosi}.  However, it is difficult to ascribe the wave to a specific
direction and/or ``source'', as for instance, the direction of
galactic rotation or the direction of the local magnetic field
\cite{kiraly}.  This is primarily because the wave is at the very
limit of MACRO statistics.  Among the possibilities for the origin of
the wave are: particle capture in magnetic traps \cite{naga}, a
close-by supernova \cite{john,erlyk}; or the motion of the solar
system though a spherical distribution of halo cosmic rays.

\begin{acknowledgments}
For J.L.M. who was there at the start and should have been here at
the end.  

We gratefully acknowledge the support of the director and
the staff of the Laboratori Nazionali del Gran Sasso and the
invaluable assistance of the technical staff of the institutions
participating in the experiment.  We thank the Istituto Nazionale di
Fisica Nucleare (INFN), the U.S. Department of Energy, and the
U.S. National Science Foundation for their generous support of the
MACRO experiment.  We thank INFN, ICTP (Trieste), WorldLab, and NATO
for providing fellowships and grants (FAI) for non-Italian citizens.
We thank Dr. U. Modigliani for helping us in obtaining and using the
temperature data from the ECMEF, Reading (UK).  We also wish to thank
the Ten. Col. Tarantino and Ispettorato Telecommunicazioni ed
Assistenza Volo dell'Aeronautica Italiana for kindly providing the
balloon temperature data.  We thank Dr. V.A. Koziarivsky of the
Institute of Nuclear Research of the Academy of Sciences, CSI, for
suggestions and discussions and Dr. K. Munakata for discussion and
exchanging information on the results of the Tibet Array.
\end{acknowledgments}

\newpage

\begin{center}
\underline {FIGURE CAPTIONS}
\end{center}

FIG. 1.  Upper panel: Hour angle distribution for the muons used in
this analysis.  The mean of the distribution falls at $-7.7^\circ$.
Lower panel: The declination distribution for the muons used in this
analysis.  The broad distribution has peaks at $\delta = 15^\circ$,
$42^\circ$, and $64^\circ$.  The mean of the distribution falls at
$\delta = 37^\circ$.

\vspace{0.25in}  

FIG. 2.  Percentage deviations of the monthly average muon rate from
 the yearly mean, $[ \Delta R(\tau_\odot)/R_0 ]_{yr}$ (solid circles),
 computed from data collected by MACRO during the years 1991-1997,
 compared with percentage deviations of the monthly average effective
 temperature from the yearly mean, $\Delta T_{eff}/<T_{eff}>$ (open
 triangles), during the same period from balloon data provided by
 Ispettorato Telecommunicazioni ed Assistenza Volo Dell'Aeronautica
 Italiana.  The computation of $T_{eff}$ is described in
 \cite{macro1}.  A small offset between the monthly means has been
 introduced for clarity.

\vspace{0.25in}  

FIG. 3.  Deviations of the muon rate from the mean muon rate binned
 according to the local solar diurnal time at the Gran Sasso.
 Superposed is the best-fit curve of the form eq.(\ref{modulation})
 representing the modulation.

\vspace{0.25in}

FIG. 4.  Superposed as open triangles onto the solar diurnal
 modulation of the mean muon rate, $\Delta R/R$, shown in Fig.~3 are
 the deviations from the daily mean of the effective temperature,
 $\Delta T_{eff}/T_{eff}$, at 0, 6, 12, and 18 hours.  The temperature
 data are taken from the European Cetre for Medium-Range Temperature
 Forcasts (ECMWF) on a grid centered on the Gran Sasso.

\vspace{0.25in}  

FIG. 5.  Deviations of the muon rate from the mean muon rate binned
 according to the local sidereal time at the Gran Sasso.  Superposed
 is the best-fit curve of the form eq.(\ref{modulation}) representing
 the modulation.

\vspace{0.25in}

FIG. 6.  Deviations of the muon rate from the mean muon rate binned
 according to the local pseudo-sidereal time at the Gran Sasso.
 Superposed is the expected pseudo-sidereal modulation from eq.(\ref{beats}).

\vspace{0.25in}

FIG. 7.  Summary of sidereal wave searches at energies $\geq 10^{12}$
 eV.  Upper panel: measurements of the amplitude; Lower panel: measurements of the phase maximum
 of the first harmonic in the sidereal variations.  Filled symbols
 represent underground experiments and open symbols represent EAS
 experiments. Symbol key: {\it large filled circle} = MACRO (this
 paper); {\it inverted filled triangle} = Poatina \cite{fenton}; {\it
 filled square} = Matsushiro \cite{mori}; {\it filled triangle} = Utah
 \cite{cutler3}; {\it asterisk} = Hong Kong \cite{lee}; {\it small
 filled circle} = Baksan \cite{andreyev}; {\it filled star} = Tibet
 \cite{muna2}; {\it open cross} = Baksan EAS \cite{alex}; {\it open
 triangle} = Kamioka \cite{muna}; {\it open square} = Norikura
 \cite{naga}; {\it open diamond} = Musala Peak \cite{gombosi}; {\it
 open circle} = EAS-TOP \cite{aglietta}; {\it open star} = Linsley
 \cite{linsley}.

\vspace{0.25in}

FIG. 8.  The sidereal wave detected by MACRO mapped into Galactic
 coordinates.  Positive wave amplitudes are shown as crosses; negative
 wave amplitudes are shown as open circles.  The wave's maximum
 amplitude, corrected for the motion of the sun with respect to the
 Local Standard of Rest \cite{mihalas} is shown as a filled star,
 assuming $\delta = 37^\circ$.  The upper filled triangle shows the
 maximum wave amplitude assuming $\delta = 64^\circ$; the lower
 filled triangle shows the maximum wave amplitude for $\delta =
 15^\circ$.  The direction toward which the sun is moving as a result
 of differential Galactic rotation is shown as a filled circle and the
 direction of the Perseus spiral arm is shown as a filled square.


\begin{table*}[t]

\begin{center}

\caption{ Results of Histogram Fits}

\begin{tabular}{|c|cccc|}
\hline
\multicolumn{1}{|c|}{Period} & 
\multicolumn{1}{c}{$K$} &
\multicolumn{1}{c}{~~~~$t_{max}$ (hr)} & 
\multicolumn{1}{c}{~~~~$\chi^2$/DoF} &  
\multicolumn{1}{c|}{~~~~$\chi^2$/DoF}\\
&&&& (null hypothesis)\\ 
\hline   
Solar Diurnal   
& $(0.88 \pm 0.26)\times 10^{-3}$ & $17.8 \pm 1.2$ & 3.4/5 & 14.6/7 \\
Apparent Sidereal  &  
$(0.82 \pm 0.27)\times 10^{-3}$ & $23.2^\dagger \pm 1.3$ & 4.6/5 & 13.8/7 \\
Pseudo-sidereal   
& $(0.0097 \pm 0.0029)\times 10^{-3}$& $4.8 \pm 2.5$ & 6.9/7 & 6.9/7 \\
\hline
\end{tabular}

\end{center}

{\small $^\dagger$ corrected for MACRO response in sidereal hour}

\end{table*}

\onecolumngrid

\begin{figure}
  \includegraphics[height=16cm,width=14cm]{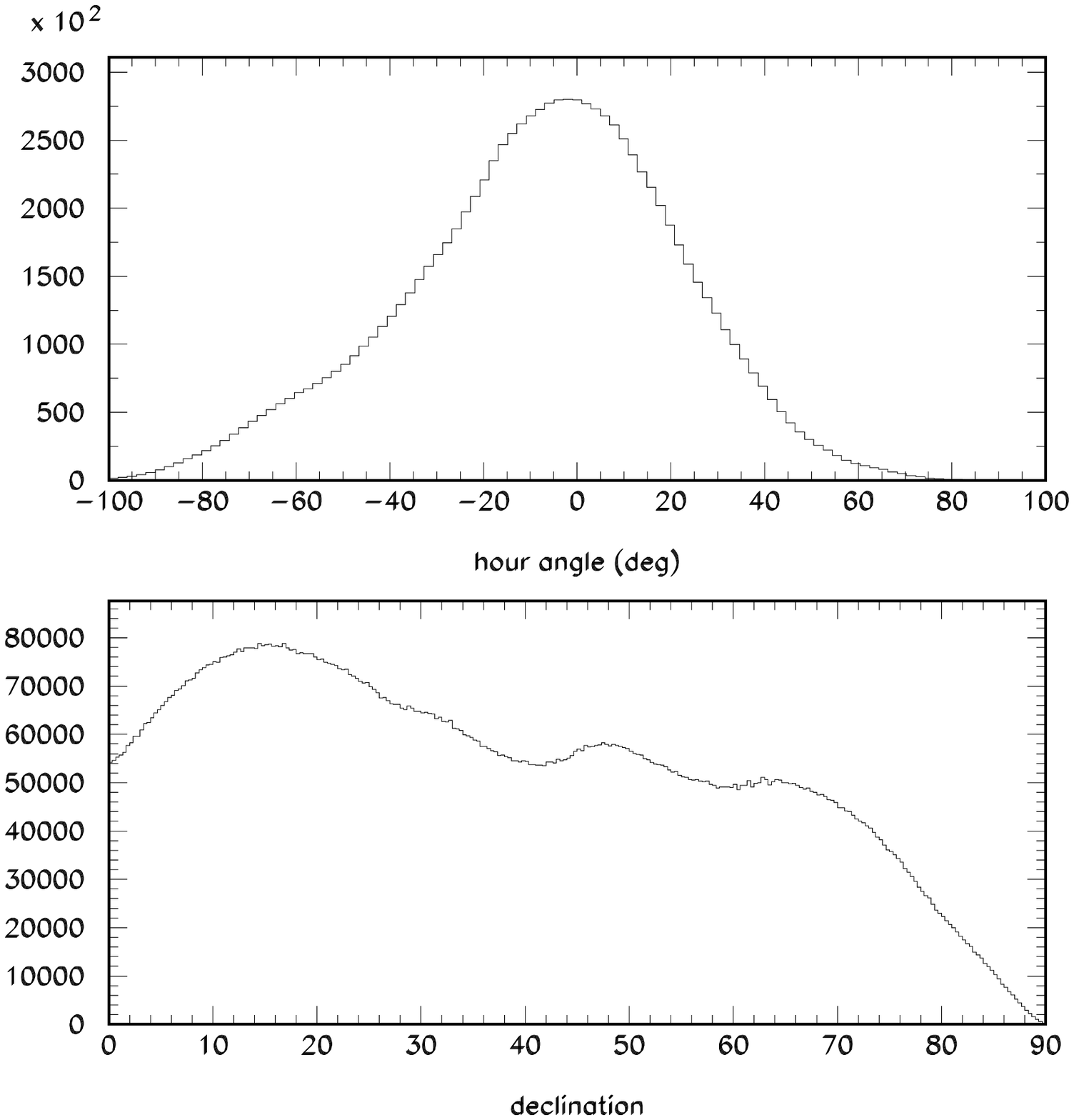}
\end{figure}

\begin{figure}
  \includegraphics[height=16cm,width=14cm]{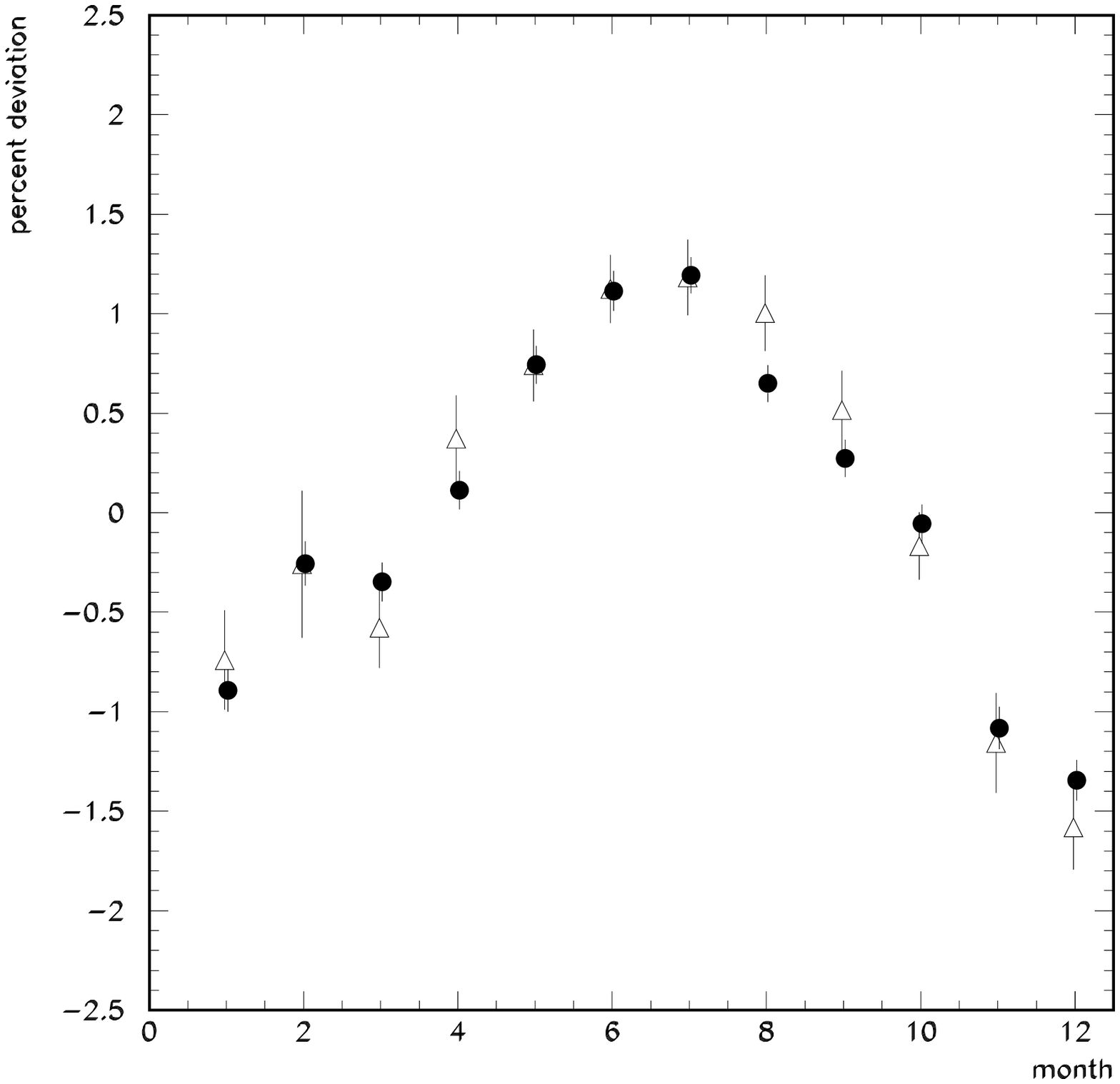}
\end{figure}

\begin{figure}
  \includegraphics[height=16cm,width=14cm]{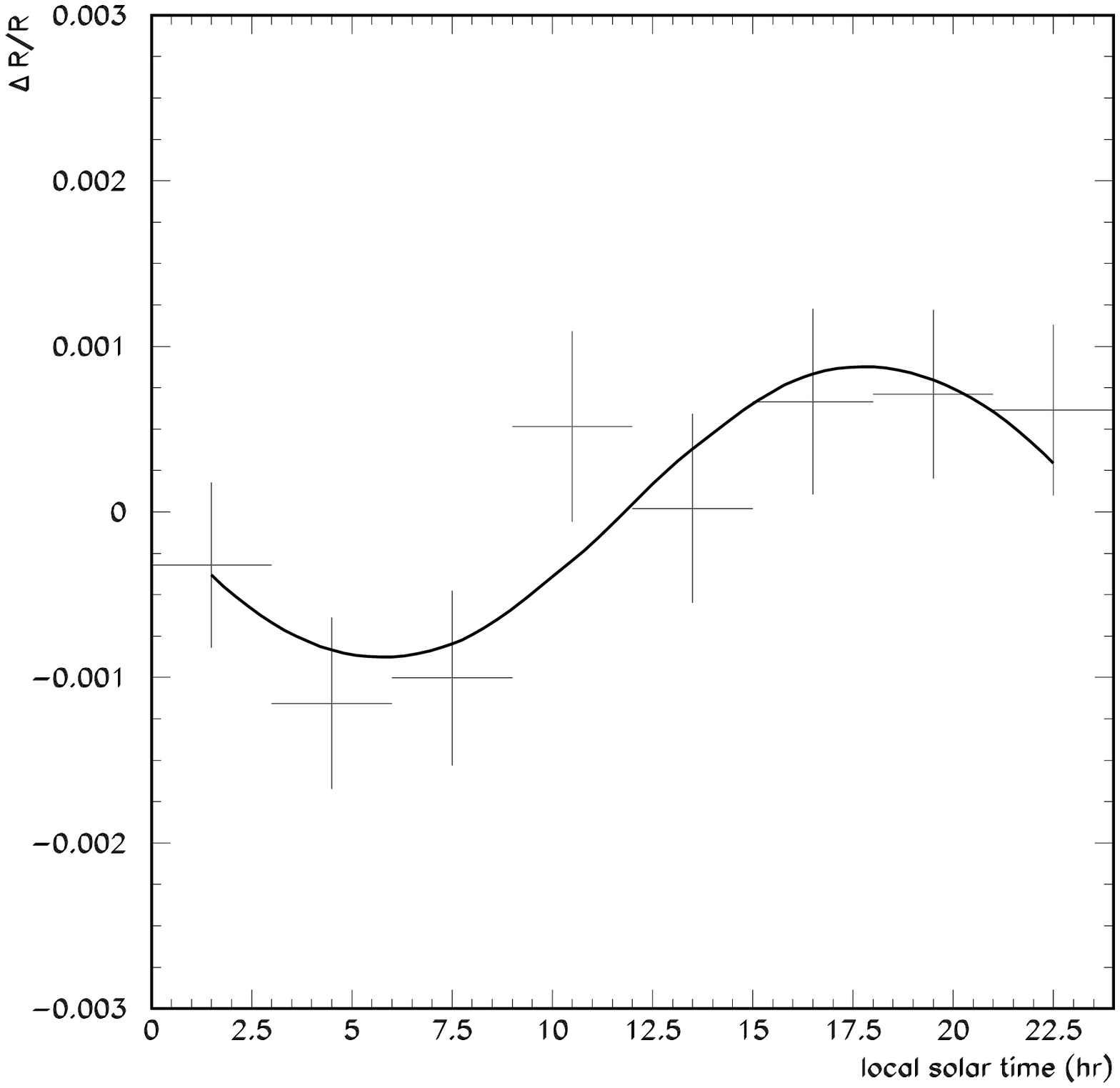}
\end{figure}

\begin{figure}
 \includegraphics[height=16cm,width=14cm]{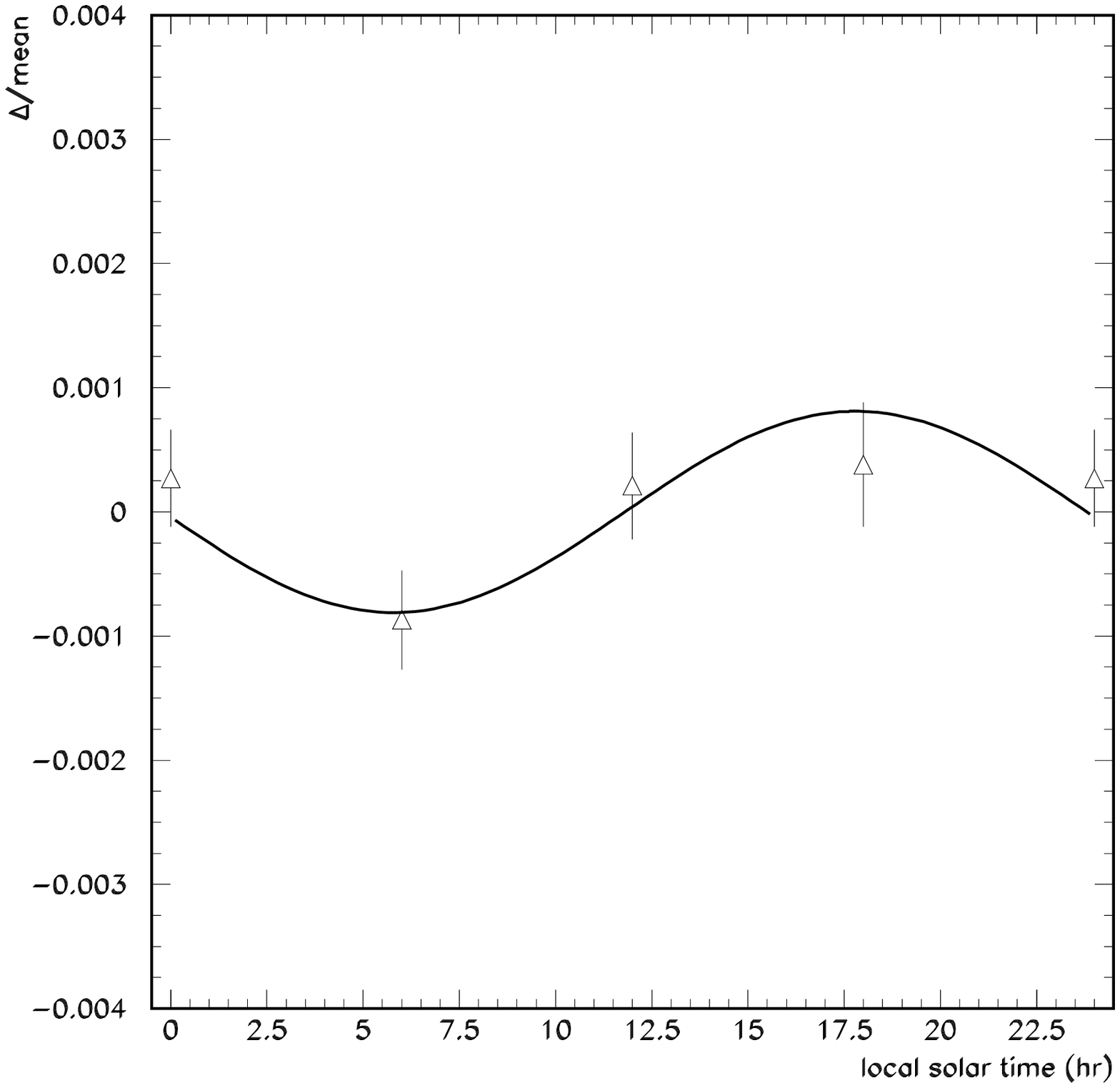}
\end{figure}

\begin{figure}
  \includegraphics[height=16cm,width=14cm]{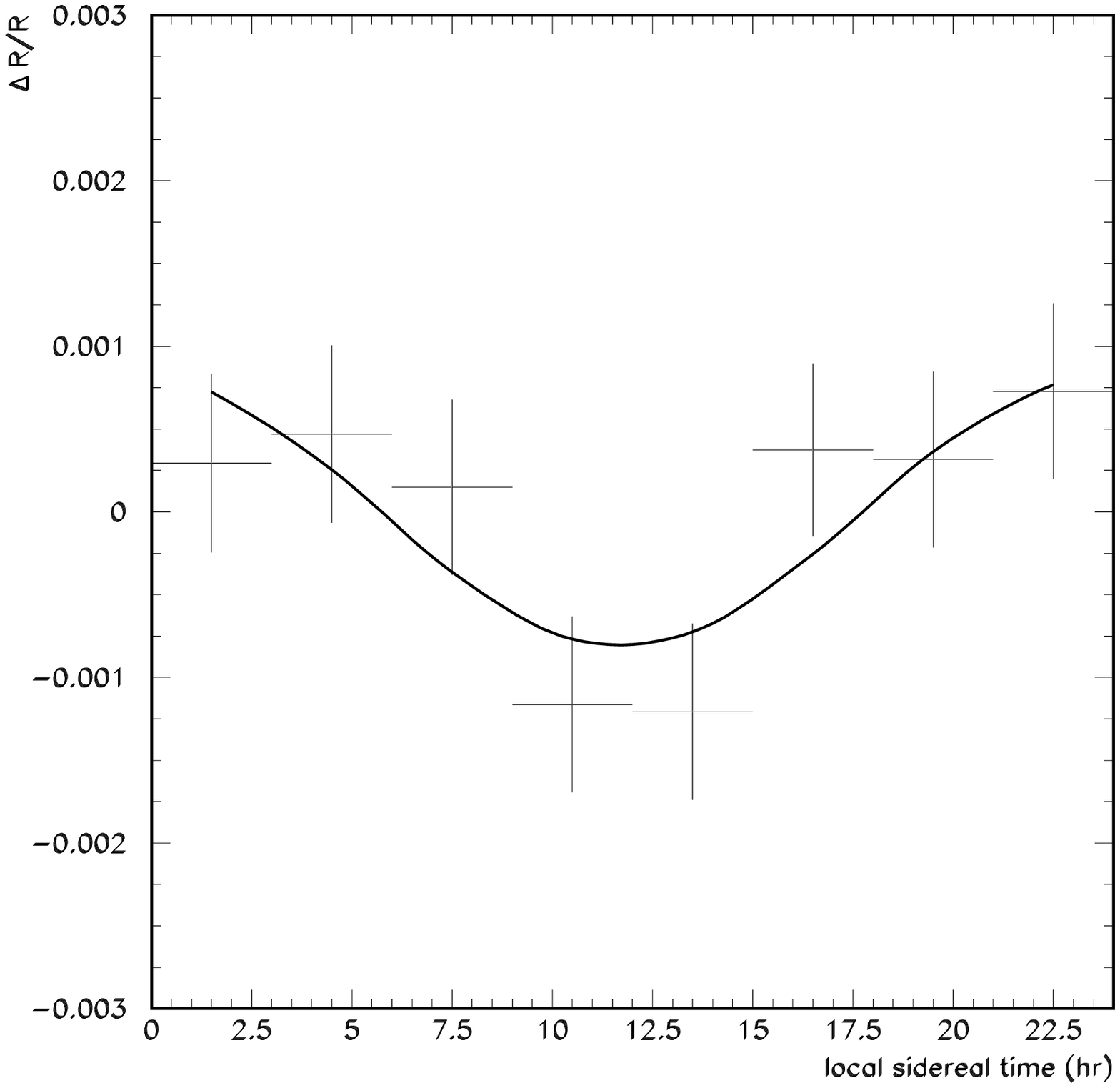}
\end{figure}

\begin{figure}
  \includegraphics[height=16cm,width=14cm]{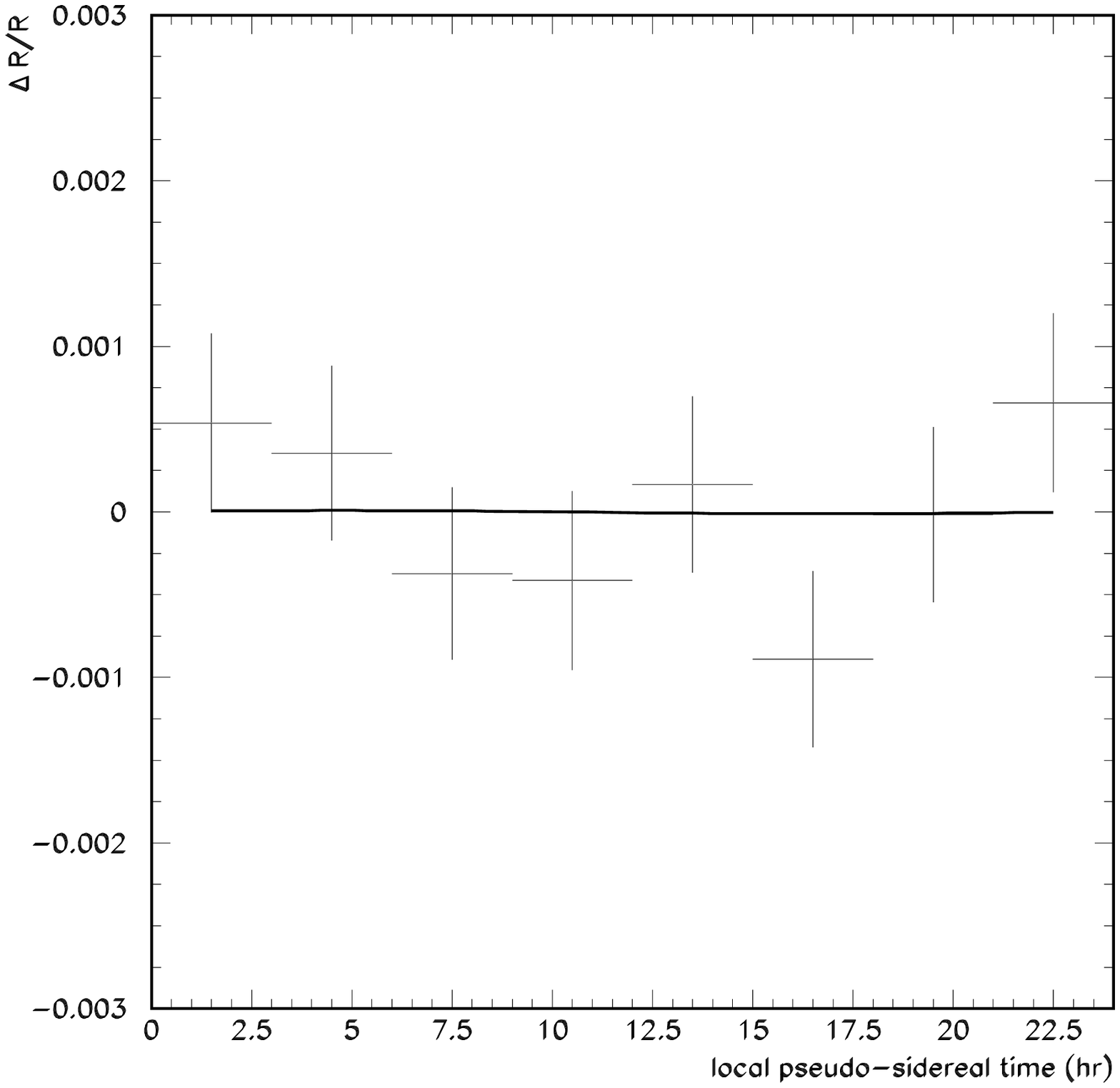}
\end{figure}

\begin{figure}
 \includegraphics[height=16cm,width=14cm]{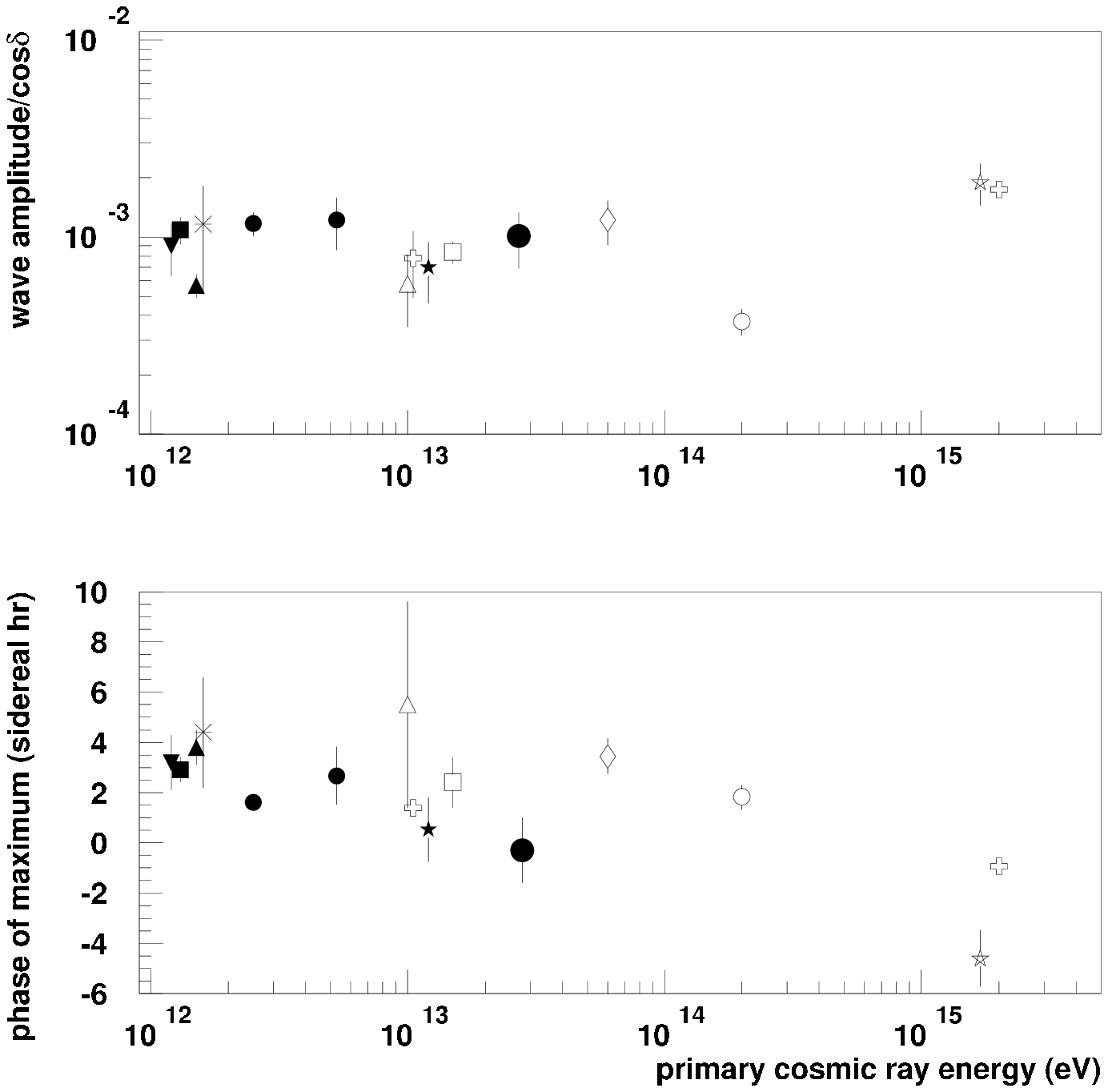}
\end{figure}

\begin{figure}
 \includegraphics[height=16cm,width=14cm]{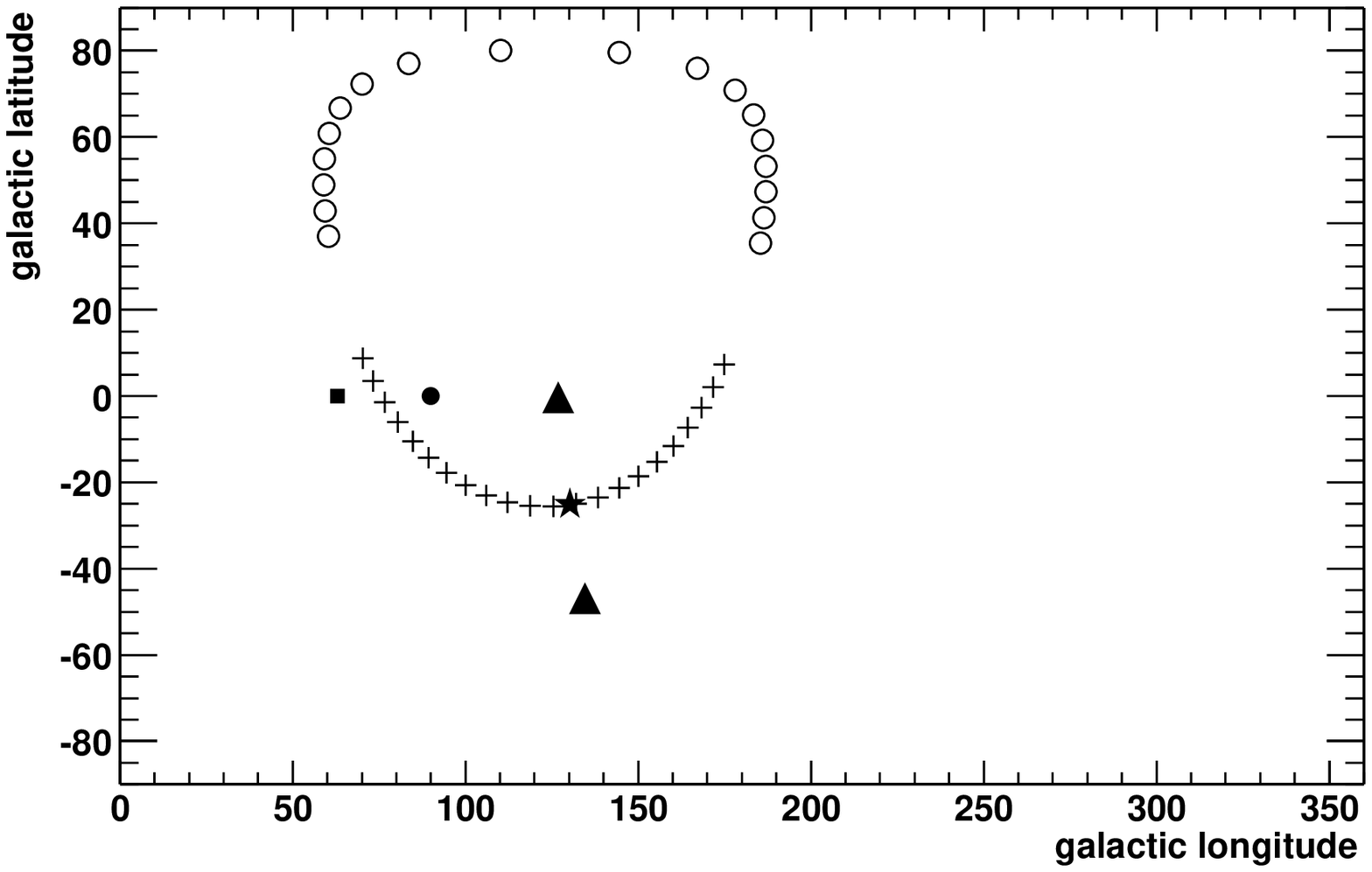}
\end{figure}


\end{document}